\documentclass[final,1p,times]{elsarticle}
\usepackage[table]{xcolor}
\usepackage{graphicx}
\usepackage{amsmath,amssymb}
\usepackage{algorithm} 
\usepackage{algpseudocode}
\usepackage{mathtools}
\usepackage{enumitem}
\usepackage{subcaption}
\usepackage{etoolbox}
\usepackage{stfloats}
\usepackage{hyperref}
\usepackage{pdfpages}
\usepackage{booktabs,longtable}
\usepackage{nomencl}
\usepackage{multicol,multirow}
\usepackage{colortbl}
\usepackage{diagbox}
\usepackage{svg}
\usepackage{amssymb}
\usepackage{nomencl}
\usepackage{vcell}
\makenomenclature
\usepackage{setspace}
\usepackage{etoolbox}
\renewcommand\nomgroup[1]{%
  \item[\bfseries
  \ifstrequal{#1}{A}{Vectors and Indices}{%
  \ifstrequal{#1}{P}{Parameters}{%
  \ifstrequal{#1}{O}{Other symbols}{}}}%
]}
\definecolor{TBLHeader}{gray}{0.85}
\definecolor{TBLRow}{gray}{0.95}
\definecolor{LightGray}{gray}{0.9}
\definecolor{Code}{rgb}{0,0,0}
\definecolor{Decorators}{rgb}{0.5,0.5,0.5}
\definecolor{Numbers}{rgb}{0.5,0,0}
\definecolor{MatchingBrackets}{rgb}{0.25,0.5,0.5}
\definecolor{Keywords}{rgb}{0,0,1}
\definecolor{self}{rgb}{0,0,0}
\definecolor{Strings}{rgb}{0,0.63,0}
\definecolor{Comments}{rgb}{0,0.63,1}
\definecolor{Backquotes}{rgb}{0,0,0}
\definecolor{Classname}{rgb}{0,0,0}
\definecolor{FunctionName}{rgb}{0,0,0}
\definecolor{Operators}{rgb}{0,0,0}
\definecolor{Background}{rgb}{0.98,0.98,0.98}
\definecolor{darkred}{rgb}{0.702, 0.529, 0.027}
\definecolor{AdGold}{rgb}{0.702, 0.529, 0.027}
\definecolor{AdRed}{rgb}{0.831, 0.047, 0.047}
\definecolor{AdBlue}{rgb}{0, 0.357, 0.612}
\definecolor{AdBlueLight}{HTML}{53E0FC}
\definecolor{AdBlueGreen}{HTML}{3DECED}
\definecolor{AdCream}{HTML}{FCFCD4}
\definecolor{AdOrange}{HTML}{ED7D31}
\definecolor{AdOrangeLight}{HTML}{FCE4D6}
\newcommand{\ceil}[1]{\lceil {#1} \rceil}

\DeclareCaptionFormat{algor}{%
  \hrulefill\par\offinterlineskip\vskip1pt%
    \textbf{#1#2}#3\offinterlineskip\hrulefill}
\DeclareCaptionStyle{algori}{singlelinecheck=off,format=algor,labelsep=space}
\usepackage{pifont}
\usepackage[switch,pagewise]{lineno}

\journal{Engineering Applications of Artificial Intelligence}

\begin{document}
\begin{frontmatter}
\title{Dynamic and Memory-efficient Shape Based Methodologies for User Type Identification in Smart Grid Applications}
%
%
\author[UoA]{Rui Yuan}
\cortext[mycorrespondingauthor]{Corresponding author}
\ead{r.yuan@adelaide.edu.au}
\author[UoA]{S. Ali Pourmousavi}
\author[UoA]{Wen L. Soong}
\author[watts]{Jon A. R. Liisberg}
\address[UoA]{The University of Adelaide, School of Electrical and Electronic Engineering, Adelaide, South Australia, Australia.}
\address[watts]{Watts A/S, Køge, Denmark.}
\begin{abstract}
Detecting behind-the-meter (BTM) equipment and major appliances at the residential level and tracking their changes in real time is important for aggregators and traditional electricity utilities. In our previous work \cite{yuan2022irmac}, we developed a systematic solution called IRMAC to identify residential users' BTM equipment and applications from their imported energy data. As a part of IRMAC, a Similarity Profile (SP) was proposed for dimensionality reduction and extracting self-join similarity from the end users' daily electricity usage data. 
The proposed SP calculation, however, was computationally expensive and required a significant amount of memory at the user's end. To realise the benefits of edge computing, 
in this paper, we propose and assess 
three computationally-efficient updating solutions, namely additive, fixed memory, and codebook-based updating methods. Extensive simulation studies are carried out using real PV users' data to evaluate the performance of the proposed methods in identifying PV users, tracking changes in real time, and examining memory usage. 
We found that the Codebook-based solution reduces more than $30\%$ of the required memory without compromising the performance of extracting users' features. When the end users' data storage and computation speed are concerned, the fixed-memory method outperforms the others. In terms of tracking the changes, different variations of the fixed-memory method show various inertia levels, making them suitable for different applications.
\end{abstract}

\begin{keyword}
\texttt{Dynamic updating, Data mining, Renewable energy, Pattern recognition, Binary classification, Time series mining, Data compression}
\end{keyword}
\end{frontmatter}





\nomenclature[1]{\(i\)}{Index for daily data}
\nomenclature[1]{\(k\)}{Index for codewords}

\nomenclature[P]{\(d_{i,j}\)}{Dissimilarity or distance between the energy time series of day $i$ and day $j$.}
\nomenclature[P]{\(d_{max}\)}{The maximum dissimilarity or distance among all pairs of the energy time series.}
\nomenclature[A]{\(\hat{d_{i}}\)}{Average dissimilarity of day $i$.}
\nomenclature[P]{\(d_{rep}\)}{The dissimilarity or distance threshold to determine whether two sub-patterns can be replaced.}
\nomenclature[P]{\(l\)}{The total length of the energy time series data.}
\nomenclature[P]{\(m\)}{The number of time intervals for one day's worth of data.}
\nomenclature[P]{\(N\)}{The number of days in the whole energy time series data.}
\nomenclature[P]{\(M\)}{The number of days of data stored in the fixed memory method.}
\nomenclature[P]{\(W\)}{The number of the codewords in codebook.}
\nomenclature[P]{\(c_{i,j}\)}{$c_{i,j} = 1$ if day $i$ and $j$ are similar. $c_{i,j} = 0$ otherwise.}
\nomenclature[P]{\(\text{TH}\)}{A similarity threshold, i.e., day $i$ and $j$ are similar if $d_{i,j}\leq \text{TH}$.}
\nomenclature[A]{\(\hat{c_{i}}\)}{The number of similar days to day $i$.}
\nomenclature[A]{\(\hat{ch_{i}}\)}{Vector representing distance changes due to the updating process.}
\nomenclature[A]{\(\text{SP}\)}{Vector of a Similarity Profile. }
\nomenclature[A]{\(\Tilde{\text{SP}}\)}{Vector of a Similarity Profile of the old time series data before updating.}
\nomenclature[A]{\(\text{SP}'\)}{Changes in the vector of Similarity Profile during updating, indexed by $i$.}

\nomenclature[A]{\(\text{TSD}\)}{Vector of metered electricity time series data by days, indexed by day $i$.}
\nomenclature[A]{\(\text{CR}\)}{Vector of the compressed representation of the whole time series data with a sequence of codewords in the codebook.}
\nomenclature[A]{\(\text{CW}\)}{Vector of codewords in the codebook, indexed by $k$.}
\printnomenclature

\section{Introduction}
\label{sec:intro}
With the increasing penetration of renewable energy resources (RES) into the modern electricity supply and demand side, the nature of supply and demand are becoming more dynamic than ever before. The International Renewable Energy Agency (IRENA) noted the total RES power capacity had doubled from $1,227$ GW to $2,537$ GW in the last decade~\cite{IRENA2020}. Due to the unpredictability of RES, modern power grid operation is threatened by frequent and rapid generation and consumption mismatch, which is likely to cause stability issues and revenue losses \cite{Kani2020}. To overcome this issue, more flexible resources are needed at different spatiotemporal scales. 

One major source of flexibility is the demand side, where information technology, automation, RES, stationary batteries and electric vehicles have turned traditional consumers into prosumers. To mobilise behind-the-meter (BTM) flexibility, the aggregator concept has been developed as an entity between prosumers and the upper grid to integrate flexible resources and motivate prosumers to adjust their electricity usage to assist network requirements~\cite{DiSomma2019}. Due to the uncertain nature of RES generation, detecting and tracking the RES owners in real-time, including accurately detecting when a RES is off or malfunctioning, are critical to aggregators, retailers, and the transmission and distribution power system operators. In some countries and jurisdictions, e.g., the USA, the retailers have access to registered Photovoltaics (PV) owners' data. However, the US national database covers only 70\% of the users because of missing data. Also, some PV owners are not eligible to get a feed-in tariff; hence are not registered at all \cite{Barbose2021}. The identification problem becomes more challenging when only the imported energy $\text{TSD}$ from users' utility meters are available, as opposed to the separate demand and RES generation data~\cite{Sun2016}.

In our previous work in~\cite{yuan2022irmac}, we tried to tackle the users' identification problem with a time series shape-based method considering the time dynamics and distortion while preserving the interpretability of our solution. The proposed approach was further used to build a binary classification method called 'IRMAC' for detecting residential smart grid applications. For instance, we applied IRMAC to the rooftop PV owners and electric heating users' identification problem respectively. The proposed approach outperformed other state-of-the-art solutions, e.g., Time-series Convolutional Neural Network (TS-CNN), and some intuitive methods such as Load Duration (LD) or yearly average profiles. 

With the increasing penetration of smart meters and BTM smart devices, we believe our Refined Motif (RM) discovery approach can be used through edge computing at the users' end. This way, only the extracted RM of each user will be communicated back to the central entity, e.g., an aggregator, to solve the classification problem. In our previous method, however, the RM method was used directly by the central entity with three months of users' consumption data to solve the identification problem. Therefore, an updating RM technique that can be performed on the users' end using edge computing tools can bring several advantages, including less central data storage requirement, lower communication bandwidth, faster computation time due to distributed computation, improved consumers' privacy preservation, lower cost and better robustness. In this paper, we propose and examine the performance of three methods for updating end users' RM dynamically. The first updating method is accumulating the new incoming data without re-calculating the Similarity Profile (SP). The second method requires a fixed size of memory to store the end user's data while the new incoming data replaces the older data in the memory. The third method is dictionary-based, where we put dissimilar patterns in a codebook for SP updating. 

To the best of our knowledge, real-time data computation at the electricity users' end has rarely been discussed in the literature. A few studies used distributed computing or federated learning that requires extensive computational power and communication bandwidth, hence are impractical for utility meters or edge computing devices~\cite{Wang2021}. We contribute to the body of knowledge by proposing three major updating methods in this paper: Additive, Fixed-memory, and codebook-based. Under the Fixed-memory method, we develop three different data-dropping strategies that offer distinct advantages and drawbacks, each of which could be useful in a different application. Under the codebook-based method, two strategies are developed that enhance data recovery and information losses. Extensive simulation studies are carried out to reveal the benefits of each method.  
The proposed solutions in this paper are fast, scalable, interpretable and flexible. We applied the proposed methods to rooftop PV users' dynamic identification to assess their performance.

This  paper  is  organised  as  follows: Section~\ref{sec:definitions} provides definitions of time series similarity measures and data compression. Section~\ref{sec:pre-work} briefly explains our previous work on IRMAC and SP, as well as the problem we are trying to solve in this paper. Section~\ref{sec:method} presents three dynamic update methods in detail. The simulation studies are reported in Section~\ref{sec:experiments}, and the results are analysed in detail. The article is concluded in Section~\ref{sec:conclution}, and future work is outlined.

\section{Definitions and Notation}
\label{sec:definitions}
This section provides descriptions of various concepts and notations used in this paper. Some concepts and notations are adopted with power system conventions in mind.
\begin{description}
\item[Definition 1.] A \emph{time series} $E\in\mathbb{R}^n$ is a sequence of real-valued numbers of $e_l\in\mathbb{R}:E = (e_1, e_2, \cdots, e_l)$, where $l$ is the length of the time series.
\end{description}

Since we need to search for local patterns instead of global ones, we must define \emph{sub-patterns}.
\begin{description}
\item[Definition 2.] \emph{Sub-pattern}: Given a time series $E = (e_1, e_2, \cdots, e_l)$, the time series partition $(e_{n_1}, e_{n_2}, \cdots, e_{n_m})$ is a sub-pattern of $E$ with a length of $m$. Since we deal with residential electricity consumption data, the length $m$ is the length of a day, and $n$ is the pattern of the $n^{th}$ day. In this paper, \emph{sub-pattern} refers to one day's worth of incoming $\text{TSD}$.
\end{description}

While searching for sub-patterns, we need a \emph{distance} metric to distinguish similar patterns from dissimilar ones. 
\begin{description}
\item[Definition 3.] \emph{Distance}: The distance $d_{i,j}$ between a pair of sub-patterns $(e_{i_1}, e_{i_2}, \cdots, e_{i_m})$ and $(e_{j_1}, e_{j_2}, \cdots, e_{j_m})$ is presented by their dissimilarity. This value can be calculated by the Euclidean distance (ED) or Dynamic Time Warping (DTW), among others. 
\end{description}
Considering the characteristics of residential electricity consumption, time shifting, and pattern distortion due to irregular consumers' behaviour, and to emphasise particular temporal patterns, we use an annotated DTW as the distance metric, as explained in our previous work~\cite{yuan2022irmac}.

We can take any time series of length $l$ and obtain the distance with each of its sub-patterns $e_{n_m}$. There are two groups of queries for retrieving similar time series by their distances, namely \emph{nearest-neighbour queries} and \emph{range queries}~\cite{Wang2008}.
\begin{description}
\item[Definition 4.] \emph{Nearest-neighbour queries}: Without assuming a threshold, the closest neighbours of the query sub-patterns can be identified, i.e., $R_{1\text{NN}_q}= \min(d(q,E_i))$ for \{$E_i, q\} \in E$.
\end{description}

Having the nearest neighbours for query sub-patterns, a \emph{Matrix Profile} can be built that leads us to find approximate similar pairs, i.e., \emph{motif}. 
\begin{description}
\item[Definition 5.] \emph{Matrix Profile (MP)}: Retrieve the closest neighbours for each pair of sub-patterns with nearest-neighbour queries, and record the distance, i.e., $\text{\textbf{MP}}[i] =R_{1\text{NN}_i} $, for $i \in [1,2,3,\cdots N]$. 
\item[Definition 6.] \emph{Motif}: It is defined as the closest sub-pattern pair of the time series; hence, the pair of patterns with the smallest value in the MP. The closest refers to two sub-patterns.
\end{description}
In this paper, however, we need a single sub-pattern repeated the most and has the minimum distance from other sub-patterns. Therefore, we need a new definition of motif, which we called \emph{Refined Motif (RM)} based on \emph{range queries}.
\begin{description}
\item[Definition 7.] \emph{Range queries}: Given a sub-pattern $q = (e_{n_1}, e_{n_2}, \cdots, e_{n_m})$, a full $\text{TSD}$ $E$ of $N$ sub-patterns $(E_i, E_2 \cdots, E_N)$, a distance measure $d$ and a threshold $\varepsilon$, we can find the set of sub-patterns $R$ in $E$ that are within the distance $\varepsilon$ from $q$, i.e., $R_{\text{RQ}_q}=
\{E_i \in E|d(q,E_i)\leq \varepsilon \}$. When $q$ is a partition of $E$, i.e., $q\in E$, range queries become part of \emph{Similarity self-join}. 
\end{description}
Having the range queries, we can create a \emph{Similarity Profile} that leads us to find the most commonly occurring partitions, i.e., \emph{refined motif} (RM).

\begin{description}
\item[Definition 8.] \emph{Similarity Profile (SP)}: By retrieving the range queries for each pair of sub-patterns and recording the number of similar patterns $c_{i}$ for each query sub-pattern, the global maximum distance $d_{\text{max}}$ and the average distance of each range query $\hat{d}_{i}$, then the similarity profile $\overrightarrow{\text{SP}}$ is a vector to the occurrences of each sub-patterns in the entire TSD. It can be presented as in Fig.~\ref{fig_SP_matrix} and Equation~\eqref{eq:similarity_profile}, where $d_{i,j}$ is the annotated DTW distance between day $i$ and day $j$, $c_{i,j}$ is a binary variable representing whether the two days are similar, the total similarity indices $\hat{c_i}$, which are integers representing the number of similar patterns with current day $i$, and the average annotated DTW distance $\hat{d_i}$ that is normalised by $\max(d_{i,j})$ to be the secondary impact factor for SP. It will differentiate the days with the same similarity indices.
\begin{figure}[!ht]
\centering
\includegraphics[width=3.1in]{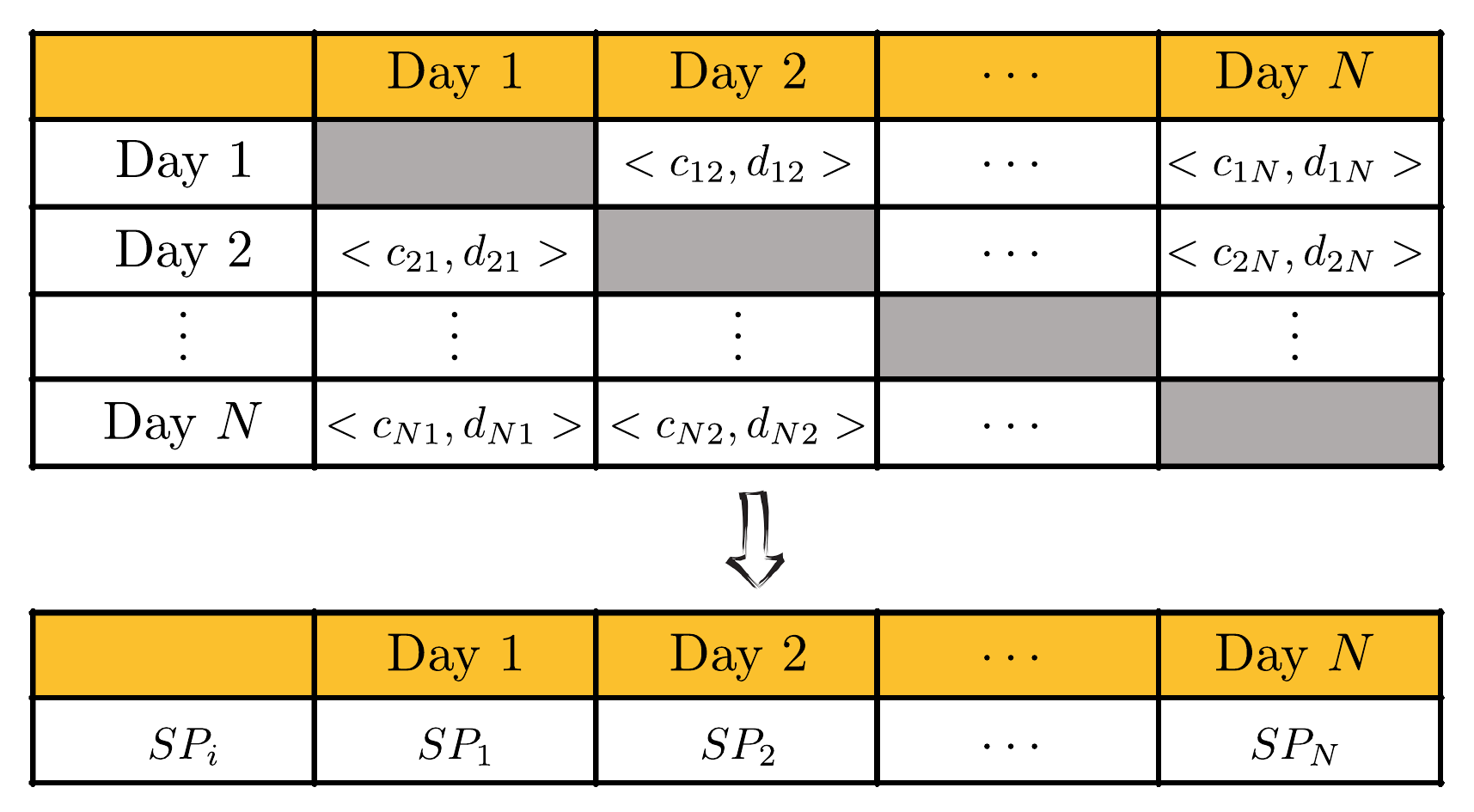}
\caption{The proposed Similarity Profile (SP) in~\cite{yuan2022irmac}}
\label{fig_SP_matrix}
\end{figure}

\begin{equation}
    \overrightarrow{SP}_{i} = \hat{c_i} - \frac{\hat{d_i}}{\max(d_{i,j})}. \label{eq:similarity_profile}
\end{equation}
\end{description}
We can retrieve the sub-pattern with the most occurrences and the least average distances from the SP, called \emph{RM}.
\begin{description}
\item[Definition 9.] \emph{RM}: It is the sub-pattern with the most similar neighbours by range queries and small global average dissimilarity. Hence, the RM is the sub-pattern with the largest value in the SP.
\end{description}
We also use the Dictionary Representation in this paper as another approach to find similarities and dissimilarities in a given time series~\cite{Wang2008, Chiarot2021}. We have two important concepts to define in the dictionary representation of a time series.
\begin{description}
\item[Definition 10.] \emph{Codeword}: A sub-pattern $\text{cw}_i$ in the time series representing a set of other similar sub-pattern neighbours whose distance is smaller than a replaceable distance threshold $d_{\text{rep}}$, i.e., $\{E_i \in E|d(\text{cw}_i,E_i)\leq d_{\text{rep}} \}$. 
\item[Definition 11.] \emph{Codebook}: It is a set of codewords $D = \{\text{cw}_1, \text{cw}_2, \cdots \text{cw}_s\}$. In Section \ref{sec: method-codebook}, we proposed a codebook dictionary approach for solving the problem.
\item[Definition 12.] \emph{Compressed representation}: For the reconstruction phase, a compressed representation $\text{cr}_i$ is the pointer that maps the compressed time series partition $i$ to the codeword $\text{cw}_s$ from the codebook, i.e., ${E_i}' = D\cdot \text{cr}_i=\text{cw}_s$.
\end{description}

\section{Related Work}
\label{sec:pre-work}
The electricity consumers' classification problem has been the focus of many research studies in the past decade~\cite{Haq2019,Butunoi2017,Bidoki2010}. However, these classification methods do not categorise consumers based on their BTM equipment. The data mining community has studied similar problems, e.g., pattern recognition and feature extraction, and various techniques have been proposed. For instance, the \emph{similarity join} method~\cite{Silva2015} has been developed and used extensively in different applications, e.g., image processing, natural language processing and databases~\cite{Zhang2021, jiang2014string, DING201920}. Later, the similarity join evolved into the \emph{similarity self-join} technique, which allows retrieving all sub-data pairs whose Euclidean distance is smaller than a predefined threshold in a given dataset~\cite{Silva2015,DeFrancisciMorales2016}. This approach has been used in time-series analysis. The similarity self-join method works based on motif discovery principles. The motif was first proposed in 2003 and defined as approximately repeated sub-patterns in a time series~\cite{chiu2003probabilistic}. Although several studies found that motifs can be used as a representative pattern in multiple applications, extracting motifs was difficult at the time as the brute-force solution was computationally untenable. In 2016, \emph{MP}, as a \emph{nearest-neighbour query} based technique, was proposed that significantly decreased the spatial and temporal complexity of the motif discovery problem by computing the z-normalised ED with the Fast Fourier Transform (FFT)~\cite{Yeh2017}. MP aims at finding the closest pair of sub-sequences; hence, it is widely used as a domain-agnostic scheme because few parameters are required \cite{Yeh2017MP4, Gharghabi2017}. 

Unlike MP, which assumes no prior knowledge of the data, we developed a highly accurate solution with low complexity for electricity consumers' binary classification with the help of our domain knowledge, e.g., considering the daily cycle of electricity usage, temporal stretch because of behavioural or seasonal changes, etc., as explained in~\cite{yuan2022irmac}. IRMAC, our previous work on the binary classification of residential smart grid applications, e.g., PV owners identification, involves two major phases. Phase one is for pattern extraction, called \emph{RM discovery}, where the sub-patterns with the highest frequency of occurrences are extracted with range queries. Based on the domain knowledge, 
we know that we need patterns during daylight hours because there is no PV generation during night hours. Additionally, to account for the consumers' rebound effect and the seasonal variations during the daylight hours, 
the sub-patterns dissimilarity is calculated using the annotated DTW method and the SP is computed as addressed in Definition 8.

The second phase of IRMAC is a classifier. It contains a single-neuron neural network with a linear function in the input layer and a sigmoid function in the output layer. This simple structure preserves the interpretability of our approach and is fast. We have previously shown in~\cite{yuan2022irmac} that IRMAC outperforms many intuitive and complex classification methods for rooftop PV users and electric heating system users' identification problems. 
Since the classifier takes RMs as input, and each user's RM discovery is independent, we proposed a structure in that RMs can be computed at the users' end, e.g., using edge computing, while only the extracted RMs will be communicated with the data centre for classification purposes. However, IRMAC needed several months of data leading to higher computation power and memory requirements at the edge of the grid. Therefore, improving IRMAC to update frequently, e.g., on a day-to-day basis, could be beneficial, as explained later in this paper. In Section~\ref{sec:method}, we propose three solutions to dynamically update RMs in an attempt to reduce memory and computational power requirements at the users' end.

\section{The Proposed RM Methods with Dynamic Update}
\label{sec:method}
\begin{figure}[!ht]
\centering
\includegraphics[clip, trim=0cm 0cm 0cm 0cm, width=.8 \textwidth]{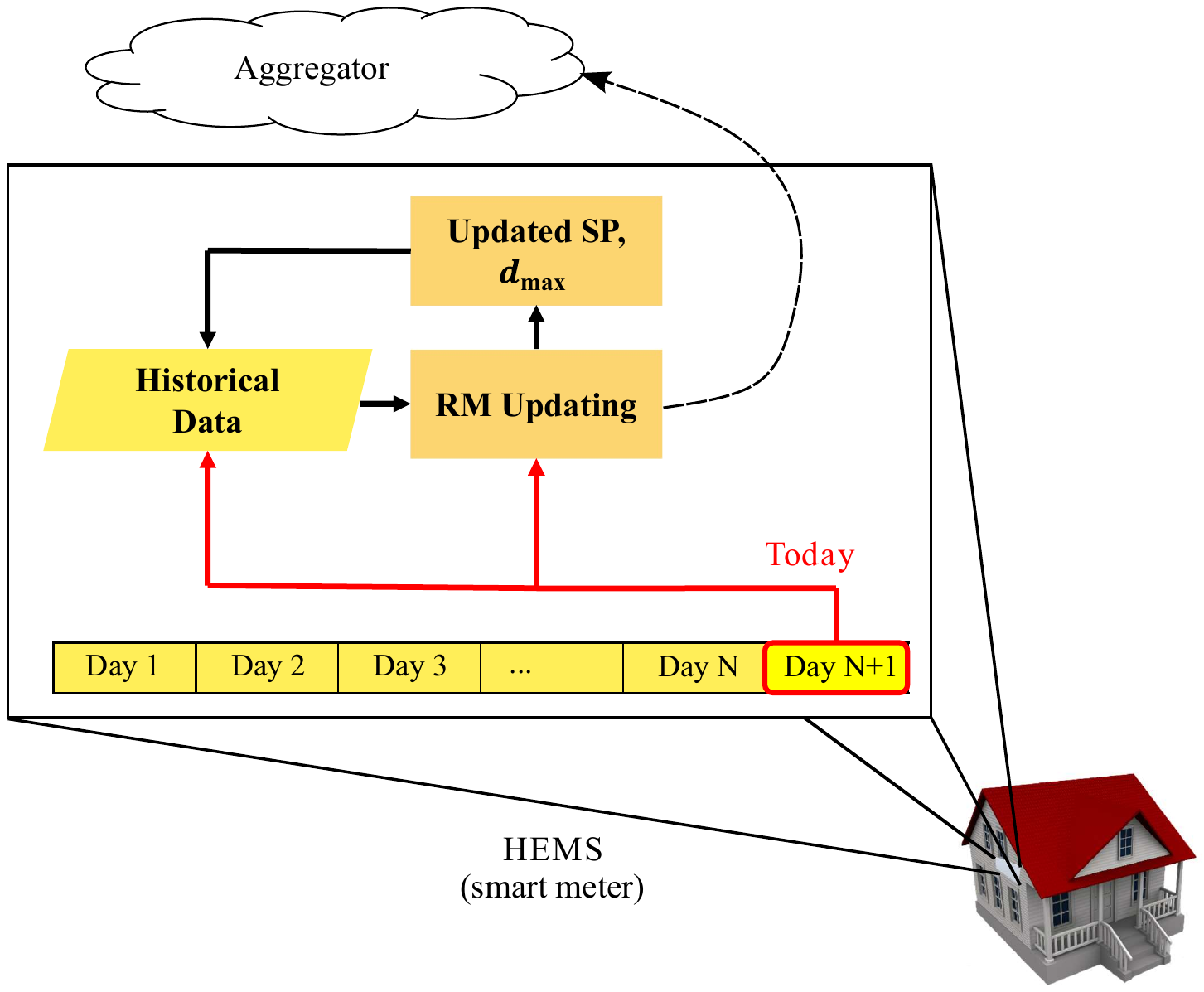}
\caption{High-level block diagram of the proposed RM updating methods}
\label{Fig:block-diagram-highlevel}
\end{figure}

The block diagram in Fig.~\ref{Fig:block-diagram-highlevel} shows a simplified and high-level framework in which the proposed dynamic update methods work. According to the existing smart meter arrangement in Australia, sub-hourly energy consumption data is fetched once at the end of each day. Therefore, our proposed updating methods will run once daily when the data is communicated. Within the home energy management system (HEMS) powered by edge computing, we will have a database of the users' SP and historical consumption data for $N$ days. The goal is to keep the end-user's SP updated upon new daily data in one iteration by the proposed RM updating methods instead of repeating the entire RM discovery process, reducing memory and computational power requirements. After updating the SP, the updated RM can be extracted from the new SP and communicated to the centre. The proposed updating solutions have different spacial complexity, inertia of tracking RM changes, and performance under different scenarios, which are discussed in the following subsections.  

The three methods are called additive, fixed memory, and codebook-based methods. Using the additive method, changes in SP are computed losslessly, although memory consumption increases over time. The fixed memory method uses a fixed memory space for updating. The codebook-based method requires a dynamic memory space. As discussed in the simulation results, the codebook-based method shows a balance between accuracy and memory-saving and outperforms the other two methods for long-term updates. The three methods are introduced in detail in the following subsections.

\subsection{RM update with Additive method}
\label{sec: method-adding}
\begin{figure}[!ht]
\centering
\includegraphics[clip, trim=2cm 4cm 8cm 3cm, width=.8 \textwidth]{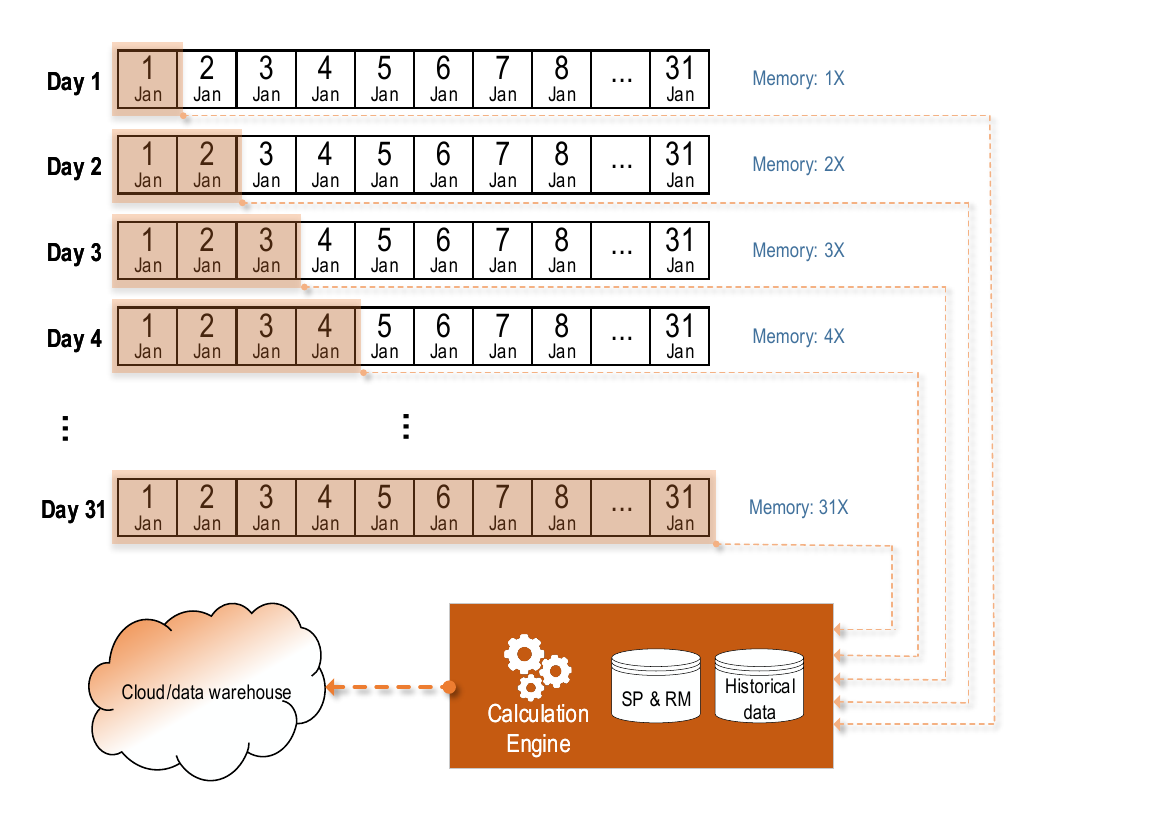}
\caption{A diagram showing the operation of the additive method and its memory requirements}
\label{Fig:block-diagram-adding}
\end{figure}

Using January as an example, the additive method can be represented schematically as
in Fig.~\ref{Fig:block-diagram-adding}. In this figure, one day's worth of $\text{TSD}$ takes the memory of $1X$ bytes, where $X$ is the memory requirement for a day. The required memory grows in a daily fashion.
In the additive method, the incoming data is compared with the existing historical patterns using annotated DTW to update the $\overrightarrow{\text{SP}}$ with the changes $\overrightarrow{\text{SP}^\prime}$. Therefore, the updated $\text{TSD}$ is the historical patterns plus the incoming daily values. After updating, the new time series and SP will be stored at the user's end, where the RM can be updated from the new SP, as shown in Fig.~\ref{fig_SP_matrix}. It can be seen in the figure that the memory requirement will grow over time. Equations~(\ref{eq:update-add-1}-\ref{eq:update-add-4}) present the additive updating process from day $N$ to day $N+1$. In the first step, the global maximum dissimilarity is updated with the distance between day $N+1$ and days $1$ to $N$. The changes in $\text{SP}$, namely ${\text{SP}^\prime_i}$, are then calculated with the new maximum dissimilarity, updated scaled averages from original SP, $\Tilde{\text{SP}_i}$, and updated counters, $c_{i,N+1}$, from Equation~(\ref{eq:update-add-2}). Lastly, the SP is updated by the modified $\Tilde{\text{SP}_i}$ with the new incoming value, $\text{SP}_{N+1}$ as in Equations~(\ref{eq:update-add-3}-\ref{eq:update-add-4}). The procedure takes one iteration to update the RM; hence, the dissimilarity calculations have a linear complexity. 
\begin{align}
    &max(d_{i,j}) = max(\Tilde{d_{max}},d_{i,N+1}),\;\; 1\leq i\leq N,\label{eq:update-add-1}\\
    &\text{SP}_{i}' = c_{i, N+1} - \frac{d_{i, N+1}\cdot \Tilde{d_{max}}}{\max(d_{i,j})}+ \frac{(\ceil{\Tilde{\text{SP}_i}} - \Tilde{\text{SP}_i}) \cdot (N-1)}{N},\;\; 1\leq i\leq N,\label{eq:update-add-2}\\
    &\text{SP}_{[1:N]} = \text{SP}_{[1:N]}'+\Tilde{\text{SP}_{[1:N]}},\label{eq:update-add-3}\\
    &\text{SP}_{N+1} = \sum_{i=1}^N c_{i,N+1}-\frac{\sum_{i=1}^N d_{i, N+1}}{\max(d_{i,j}) \cdot N}.\label{eq:update-add-4}
\end{align}

Algorithm~\ref{update:adding} shows a step-by-step computation process. For a $\text{TSD}$ with length $l$ and window size $m$, which is growing daily, recomputing the SP takes $O(l^2m)$ time complexity, as discussed in ~\cite{yuan2022irmac}. On the other hand, updating an incoming day's data at the user's end with the additive method on the same data length takes the time complexity of $O(l/m\cdot m^2) = O(lm)$.
\begin{center}
  \captionof{algorithm}{Additive Method for Updating SP} 
  \label{update:adding}
	\begin{algorithmic}[]
	    \Require {$\text{TSD}$ contains $N$ days data $\text{TSD}_{[1:N]}$ with a new incoming data $\text{TSD}_{N+1}$ to be updated, $\text{SP}$ with $N$ days' record $\text{SP}_{[1:N]}$, a maximum dis-similarity $d_{max}$, and a threshold $\text{TH}$}
	    \State $c_{[1:N]} \leftarrow [0,0,\dots 0]$\Comment{initialise counters}
	    \State $ch_{[1:N]} \leftarrow [0,0,\dots 0]$\Comment{initialise changes of distances}
	    \State $\hat{d}_{N+1} \leftarrow 0$\Comment{initialise dissimilarity for new data}
		\For {$i=1,2\ldots N$}

		    \State $d_{i,{N+1}} \leftarrow DTW(\text{TSD}_{N+1}, \text{TSD}_{i})$
		    \State $\hat{d}_{N+1} \leftarrow \hat{d}_{N+1}+ d_{i,{N+1}}$
		    \State ${d}'_{max} \leftarrow \max(d_{max}, d_{i,{N+1}})$\Comment{update max distance}
		    \State $ch_{i} \leftarrow d_{i,{N+1}}$
			\If {$d_{i,{N+1}}\leq \text{TH}$}\Comment{check threshold}
			    \State $c_{i} \leftarrow c_{i} + 1$
			    \State $c_{N+1} \leftarrow c_{N+1}+1$
			\EndIf
		\EndFor
		\State $avg_{scaled} \leftarrow \ceil{\text{SP}_{[1:N]}}-\text{SP}_{[1:N]}  $
		\State $avg' \leftarrow \frac{avg_{scaled}\cdot(N-1)\cdot d_{max}+ch_{[1:N]}}{d'_{max}\cdot N}-avg_{scaled}  $ 
		\State $\text{SP}_{[1:N]} \leftarrow \text{SP}_{[1:N]}+c_{[1:N]}-avg' $\Comment{Equation~(\ref{eq:update-add-2}-\ref{eq:update-add-3})}
		\State $\text{SP}_{N+1} \leftarrow c
		_{N+1} - \frac{d_{new}\cdot (N+1)}{d'_{max}\cdot N}$\Comment{Equation~(\ref{eq:update-add-4})}
		\State \Return $\text{SP}$, $\text{TSD}$, $d'_{max}$\Comment{RM is $\text{TSD}_{i}$ where $\text{SP}_{i} == \text{max}(\text{SP})$}
	\end{algorithmic}
\noindent\makebox[\linewidth]{\rule{0.4 \paperwidth}{0.4pt}}
\end{center}

The main bottleneck of the additive method is that the memory and computational time requirements increase with the number of days, as shown in Fig.~\ref{Fig:block-diagram-adding}. In addition, the additive method provides a global RM by considering the entire historical data. Therefore, it can be as accurate as the proposed method in~\cite{yuan2022irmac}. However, this method is less sensitive to identifying the user's type switching, i.e., a non-PV user switching to a PV user or vice versa. To address these issues, we propose a modified updating approach in the next subsection.  

\subsection{RM update with Fixed-memory method}
\label{sec: method-fix}
\begin{figure}[!ht]
\centering
\includegraphics[clip, trim=2cm 2cm 8cm 3cm, width=.8 \textwidth]{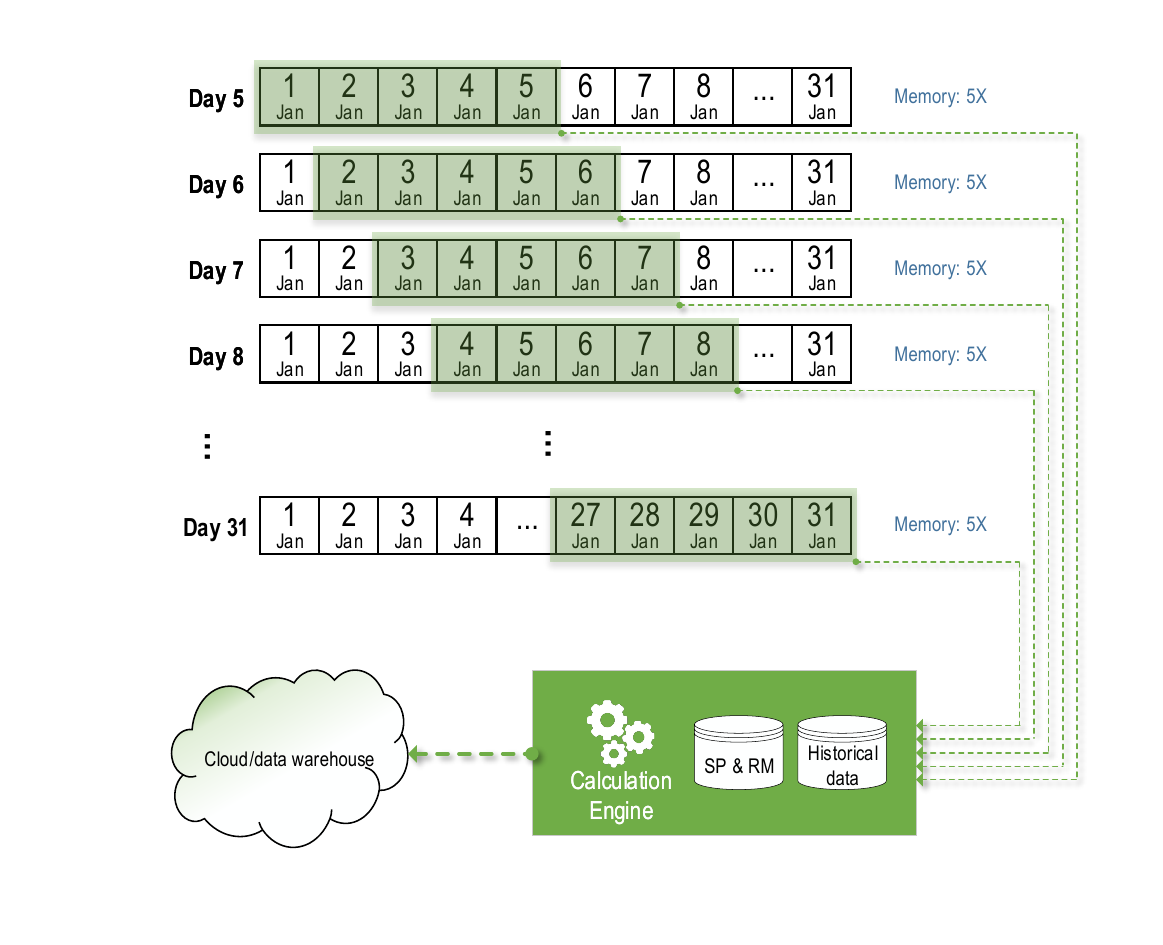}
\caption{A framework representing the operation of the Fixed-memory method and its memory requirement}
\label{Fig:block-diagram-fix}
\end{figure}

This method uses a fixed memory of historical values and SP for updating, as shown in Fig.~\ref{Fig:block-diagram-fix}. Consequently, each new incoming sub-pattern data replaces one sub-pattern in the memory. In other words, one day's worth of sub-pattern needs to be dropped from the memory to maintain the fixed size of the memory requirement. Hence both updated SP and time series preserve the same length and memory, which is set by the size of the allocated memory, $M$. Compared to the additive method, the time complexity of the fixed-memory method is $O(M\cdot m)$, which is $l/M$ times the complexity as $l$ is normally much larger than $M$. The step-by-step process of the fixed-memory method is presented in Algorithm~\ref{update:fix}. To select which data to drop, we propose three approaches in this study, which are explained later in this section. 

\begin{center}
  \captionof{algorithm}{Fixed-memory method} 
  \label{update:fix}
	\begin{algorithmic}[]
	    \Require {variable $\text{TSD}$ contains $M$ stored daily sub-patterns $\text{TSD}_{[1:M]}$ with an incoming day's data $\text{TSD}_{M+1}$ to update, variable $\text{SP}$ with M days' similarity information $\text{SP}_{[1:M]}$, a maximum dis-similarity $d_{max}$, and a threshold $\text{TH}$}
	    \State $\text{TSD}_{in} \leftarrow \text{TSD}_{M+1}$\Comment{new incoming data}
	    \State $\text{TSD}_{out} \leftarrow \text{TSD}_{index}$\Comment{drop data by index}
	    \State Remove $\text{TSD}_{index}$ from $\text{TSD}$
	    \State $c_{[1:M]} \leftarrow [0,0,\dots 0]$\Comment{initialise counters}
	    \State $ch_{[1:M]} \leftarrow [0,0,\dots 0]$\Comment{initialise changes of distances}
	    \State $d_{new} \leftarrow 0$\Comment{initialise dissimilarity for new data}
		\For {$i=1,2,3\ldots \text{M-1}$}
		    
		    \State $d_{out} \leftarrow \text{DTW}(\text{TSD}_{out}, \text{TSD}_{[i]})$
		    \State $d_{in} \leftarrow \text{DTW}(\text{TSD}_{in}, \text{TSD}_{[i]})$
		    \State $d_{new} \leftarrow d_{new}+ d_{in}$
		    \State $d'_{max} \leftarrow \max(d_{max}, d_{in})$\Comment{update distance}
		    \State $ch_{i} \leftarrow d_{in}-d_{out}$
		    \State $ch_{i} \leftarrow \frac{ch_{i}}{M}$\Comment{changes in average}
			\If {$d_{out}\leq TH$}\Comment{check threshold}
			    \State $c_{[i]} \leftarrow c_{[i]} - 1$
			\EndIf
			\If {$d_{in}\leq TH$}\Comment{check threshold}
			    \State $c_{i} \leftarrow c_{i} + 1$
			    \State $c_{M} \leftarrow c_{M+1}+1$
			\EndIf
		\EndFor
		\State $avg_{scaled} \leftarrow \ceil{\text{SP}_{[1:M-1]}}-\text{SP}_{[1:M-1]}  $
		\State $avg' \leftarrow \frac{avg_{scaled}\cdot d_{max}+ch[1:M-1]}{d'_{max}}-avg_{scaled} $
		\State $\text{SP}_{[1:M-1]} \leftarrow \text{SP}_{[1:M-1]}+c_{[1:M-1]}-avg'$
		\State $\text{SP}_{M} \leftarrow c_{M} - \frac{d_{new}}{d'_{max}}$
		\State \Return $\text{SP}$, $\text{TSD}$, $d'_{max}$\Comment{RM is $\text{TSD}_{i}$ where $\text{SP}_{i} = max(\text{SP})$}
	\end{algorithmic}
\noindent\makebox[\linewidth]{\rule{0.4 \paperwidth}{0.4pt}}
\end{center}

This method is a lossy solution in terms of preserving the actual SP values due to the method tracking the maximum distance within the memory rather than the global maximum distance, i.e., $\text{max}(d_{i,j})$ in Equation~\eqref{eq:update-add-1}. The extracted RM remains accurate as the decimal relationship in the SP is preserved, with its decimal part scaled at the same level in Equation~\eqref{eq:update-add-4}. Two additional considerations for the Fixed-memory method, i.e., sub-pattern dropping strategy and the window size, are discussed next.

\subsubsection{Different dropping strategies}
\label{sec:method-fix-drop}
Since the required memory in this method is fixed, we need to drop one day's worth of data from memory upon receiving each new daily data. Three dropping strategies are proposed, which could be used in different situations for different applications that are discussed below:

\begin{enumerate}
    \item \textbf{Low-inertia strategy}: In this approach, we eliminate the oldest day's data from the memory to save the new incoming daily data. This way, we can access the newest $M$ days of consumer demand profiles. As a result, this strategy is heavily influenced by the most recent historical data, and the temporal dynamics of this data will play the main role in finding the RM. In other words, the RMs can change quickly with changes in the consumers' behaviour; hence, is a low-inertia or short-memory strategy. 
    \item \textbf{High-inertia strategy}: Since the daily $\text{TSD}$ and SP are stored and do not need to be consecutive, we propose dropping the data with the lowest SP value to preserve some long-term memory by introducing a ``high-inertia strategy''. The intuitive explanation is that the sub-pattern with the highest dissimilarity indicates an abnormality that can be ignored. This strategy will preserve similar patterns in the memory and enhance the featured patterns. Therefore, it can discover more accurate RMs for most users. However, the main drawback of this method is that the stored patterns are relatively similar; hence, they lack variety. This will make it more difficult to identify changes in the consumers' behaviour and type. Therefore, this strategy offers the highest inertia among the three variations.  
    \item \textbf{Medium-inertia strategy}: Considering the issues from the previous dropping strategies, we propose the third strategy in which we eliminate the $\text{TSD}$ with the median SP value. The idea is to keep the most similar and dissimilar patterns in the database. This way, the medium-inertia strategy better identifies changing consumers' behaviour compared to the high-inertia strategy while preserving the most significant behavioural patterns compared to the low-inertia approach.
\end{enumerate}

\begin{figure*}[!ht]
    \centering
    \includegraphics[clip, trim=0.1cm 8cm .1cm .1cm,width=1 \textwidth]{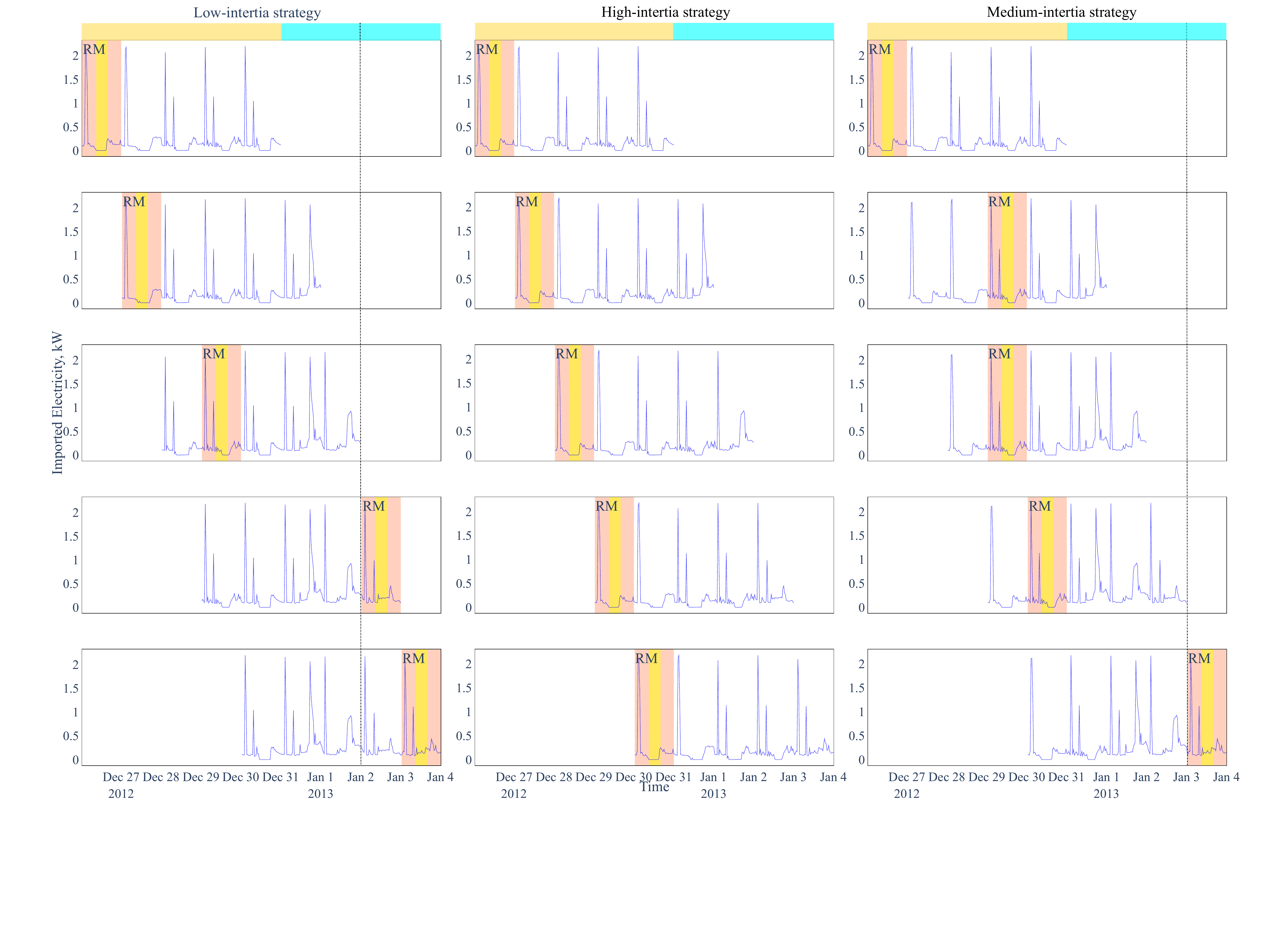}
    \caption{Three variations of the Fixed-memory method (namely, low-inertia, high-inertia and medium-inertia strategy), comparison with window size 5 detecting solar type switching}
    \label{fig:nonsolar-solar}
\end{figure*}

To better explain the three strategies, let's see an example in which a PV solar system of a consumer breaks down; hence, temporarily switching from a solar user to a non-solar user, as shown in Fig.~\ref{fig:nonsolar-solar}. The yellow and light blue areas in the figure present the periods before and after the solar system breakdown respectively. It can be seen that it takes three and four updates for the low-inertia and medium-inertia strategies, respectively, to identify the breakdown (shown by vertical line). On the other hand, the high-inertia strategy cannot detect this change even after four updates. Therefore, it can be concluded that the low-inertia strategy method is good at tracking even the smallest changes in user behaviour. In contrast, the high-inertia strategy works best for users with stable behaviour. A comprehensive simulation study on the capabilities of the three methods is presented in Section~\ref{sec:experiments}.
\subsubsection{Window size and other considerations}
The fixed-memory approach involves a new parameter, i.e., $M$, which determines how many days of sub-patterns needs to be preserved in the memory. Intuitively, the bigger $M$ means more data can be preserved throughout the updating process. However, it comes at a higher cost of memory and computation. The size of $M$ in this particular application for PV users identification depends on the consumers' behavioural changes, weather regime in an area, etc. For instance, consumers with higher behavioural changes or in an area with rapid changes in weather regimes require a higher $M$. We discussed the optimal window size in Section~\ref{sec:experiments} through simulation studies.

\subsection{RM update with Codebook-based method}
\label{sec: method-codebook}

\begin{figure}[!ht]
\centering
\includegraphics[clip, trim=4cm 1.8cm 5cm 3cm, width=.8 \textwidth]{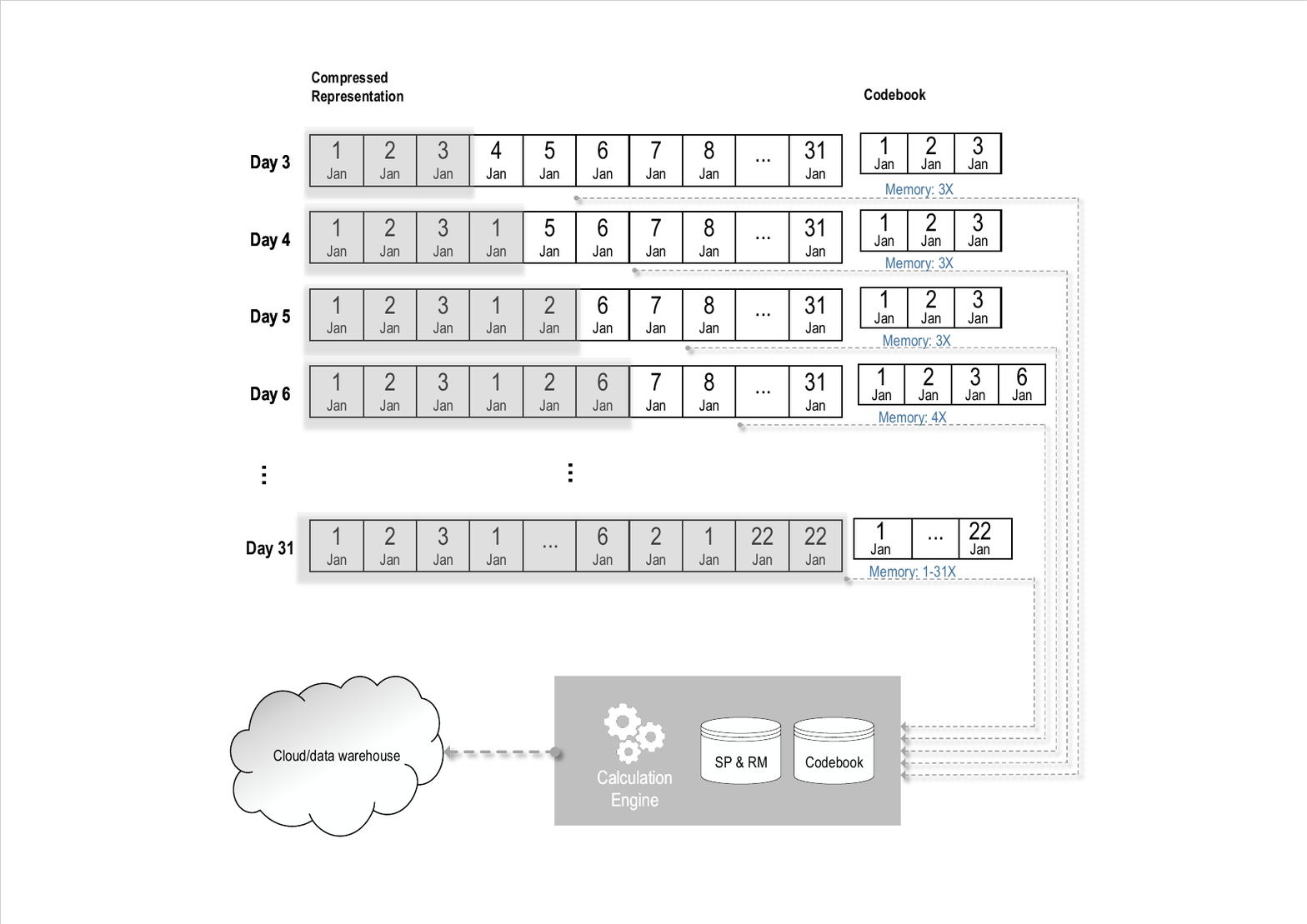}
\caption{The framework of the Codebook method for dynamic RM discovery}
\label{fig:bd-codebook}
\end{figure}

Dictionary-based methods, especially codebook models, have been used for time series compression for a long time~\cite{Wang2008, Chiarot2021}. In this paper, we adopt the same idea to the real-time updating problem of RM discovery, as shown in the block diagram of Fig.~\ref{fig:bd-codebook}. Since the dissimilarity of the daily patterns in our proposed dynamic RM is measured independently, the RM discovery process does not require the sub-patterns to be in a temporal sequence. Consequently, we propose two sub-strategies for the Codebook method that are discussed in the following subsections. 

\subsubsection{Codebook method with compressed representation}
The first strategy is the conventional codebook method. In this method, a set of sub-patterns, i.e., codebook, will be stored at the users' end containing irreplaceable patterns, called codewords, and the historical SP. The distance between the new incoming data and the existing codewords in the codebook is computed. If the new incoming data is related to the most similar codeword with a distance smaller than a pre-defined threshold, it will be replaced with an existing codeword. Otherwise, it will be added to the codebook as a new codeword. To recover the historical $\text{TSD}$ from the codewords, the end users need to store a compressed representation, i.e., a set of pointers mapping the codewords into the $\text{TSD}$ of each day. 

The codebook, i.e., the compressed representation of $\text{TSD}$ and the SP, can be updated using the daily incoming data. In data compression applications, codebook models are lossy representations of the original signal. The first method will compress the end-users data with a codebook and recover the data with compressed representation before the updating process. The updating pseudo code for this approach is presented in Algorithm~\ref{alg:codebook-1}.  

\begin{center}
  \captionof{algorithm}{Codebook with compressed representation} 
  \label{alg:codebook-1}
	\begin{algorithmic}[]
	    \Require {variable $\text{CW}$ contains $W$ stored codewords $\text{CW}_{[1:W]}$ with a incoming $\text{TSD}_{in}$ to update, variable $\text{SP}$ with $N$ days' similarity $\text{SP}_{[1:N]}$, a compressed representation with $N$ day's code number $\text{CR}_{[1:N+1]}$, a maximum dis-similarity $d_{max}$, and a distance value which is seen as replaceable $d_{rep}$}
	    \State $c_{[1:N]} \leftarrow [0,0,\dots 0]$\Comment{initialise counters}
	    \State $ch_{[1:N]} \leftarrow [0,0,\dots 0]$\Comment{initialise changes of distances}
	    \State $d_{new} \leftarrow 0$\Comment{initialise dissimilarity for the new data}
	    \State $d_{min} \leftarrow d_{rep}$
	    \State $\text{TSD}_{[1:N]} \leftarrow \text{CW}_{\text{CR}_{[1:N]}}$\Comment{recover TSD}
	    \State $\text{CR}_{N+1} \leftarrow W+1$\Comment{initialise codebook}
		\For {$i=1,2,3\ldots N$}
		    \State $d_{in} \leftarrow DTW(\text{TSD}_{in}, \text{TSD}_i)$
		    \State $d_{new} \leftarrow d_{new}+ d_{in}$
		    \State $d'_{max} \leftarrow \max(d_{max}, d_{in})$\Comment{update $d$}
		    \State $ch_{[i]} \leftarrow d_{in}$
			\If {$d_{in}\leq TH$}\Comment{check threshold}
			    \State $c_{i} \leftarrow c_{i} + 1$
			    \State $c_{N+1} \leftarrow c_{N+1} + 1$
			\EndIf
			\If {$d_{in}\leq d_{min}$}\Comment{check if replaceable}
			    \State $d_{min} \leftarrow d_{in}$
			    \State $\text{CR}_{N+1} \leftarrow \text{CR}_{i}$
			\EndIf
		\EndFor
		\State $avg_{scaled} \leftarrow \ceil{\text{SP}_{[1:N]}}-\text{SP}_{[1:N]}  $
		\State $avg' \leftarrow \frac{avg_{scaled}\cdot (N-1) \cdot d_{max}+ch[1:N]}{d'_{max}\cdot N}-avg_{scaled} $
		\State $\text{SP}_{[1:N]} \leftarrow \text{SP}_{[1:N]}+c_{[1:N]}-avg'$
		\State $\text{SP}_{N+1} \leftarrow c_{N+1} - \frac{d_{new}}{d'_{max}}$
		\If {$\text{CB}_{N+1} == W+1$}
		\State $\text{CR}_{W+1} \leftarrow \text{TSD}_{in}$
		\EndIf
		\State \Return $\text{SP}$, $\text{CW}$, $d'_{max}$, $\text{CR}$\Comment{RM is $\text{CW}_{\text{CR}_i}$ where $\text{SP}_{i} == \text{max}(\text{SP})$}
	\end{algorithmic}
\noindent\makebox[\linewidth]{\rule{0.4 \paperwidth}{0.4pt}}
\end{center}

\subsubsection{Codebook without compressed representation}
Compared to the previous codebook strategy, the second strategy does not need a compressed representation to recover the $\text{TSD}$ from a sequence of codewords. 
Instead, a new variation of the codebook, Patterns Dictionary (PD), will be stored on the end-user's side. PD is proposed to store the codewords as keys and occurrences as values. From the codewords and occurrences, the end user can store and update the entire SP, but not in a temporal fashion as the index of each value in the SP does not present the time when it is collected. Compared to the Codebook model with compressed representation, this strategy preserves the relationship of each sub-pattern but in an anonymous way because the temporal sequences are not included. More specifically, the SP's sub-pattern with a higher index does not necessarily refer to newer data than the sub-pattern with a lower index. Furthermore, it requires less memory since we do not need a compressed representation. The implementation of this strategy is explained step-by-step in Algorithm~\ref{PD-withoutcode}. 
\begin{center}
  \captionof{algorithm}{Codebook method without compressed representation} 
  \label{PD-withoutcode}
	\begin{algorithmic}[]
	    \Require {a dictionary with $W$ stored daily patterns as keys and occurrences as values, i.e. $\text{PD}=\{\text{CR}_1:n_1,...\text{CR}_W:n_W\}$, an incoming $\text{TSD}_{in}$ to update, variable $\text{SP}$ with $N$ days $\text{SP}_{[1:N]}$, a maximum dis-similarity $d_{max}$, and a value of distance which can be seen as pre-defined $d_{rep}$}
	    \State $c_{[1:N+1]} \leftarrow [0,0,\dots 0]$\Comment{initialise counters}
	    \State $ch_{[1:N]} \leftarrow [0,0,\dots 0]$\Comment{initialise changes of distances}
	    \State $d_{new} \leftarrow 0$\Comment{initialise dissimilarity for new data}
	    \State $d_{min} \leftarrow d_{rep}$
	    \State $p \leftarrow W+1$\Comment{initialise the position for new data in PD}
	    \State $cr \leftarrow 1$\Comment{track the position in SP}
		\For {$i=1,2,3\ldots W$}
		    \State $d_{in} \leftarrow DTW(\text{TSD}_{in}, \text{CR}_{i})$
		    \State $n_i \leftarrow \text{PD}_{\text{CR}_{i}} $\Comment{occurrence of the data}
		    \State $d_{new} \leftarrow d_{new}+ d_{in}\cdot n_i$
		    \State $d'_{max} \leftarrow \max(d_{max}, d_{in})$\Comment{update $d$}
		    \State $ch_{[cr:cr+n_i]} \leftarrow  {d_{in}}$
			\If {$d_{in}\leq TH$}\Comment{check threshold}
			    \State $c_{[cr:cr+n_i]} \leftarrow c_{[cr:cr+n_i]} + 1$
			    \State $c_{N+1} \leftarrow c_{N+1}+1$
			\EndIf
			\If {$d_{in}\leq d_{min}$}\Comment{check if replaceable}
			    \State $d_{min} \leftarrow d_{in}$
			    \State $p \leftarrow i$
			\EndIf
			\State $cr \leftarrow  cr+n_i$ \Comment{update the position}
		\EndFor
		\State $avg_{scaled} \leftarrow \ceil{\text{SP}_{[1:N]}}-\text{SP}_{[1:N]}  $
		\State $avg' \leftarrow \frac{avg_{scaled}\cdot (N-1) \cdot d_{max}+ch[1:N]}{d'_{max}\cdot N}-avg_{scaled} $
		\State $\text{SP}_{[1:N]} \leftarrow \text{SP}_{[1:N]}+c_{[1:N]}-avg'$
		\State $\text{SP}_{in} \leftarrow c_{[N+1]} - \frac{d_{new}}{d'_{max}}$\Comment{new data's SP}
		\If {$p \leq W+1$}
    		\State $n_{p} \leftarrow n_{p}+1$
    		\State $cr \leftarrow n_1+n_2+\ldots n_{p}$\Comment{find the position to insert new SP}
    		\State $\text{SP}_{[cr+1:N+1]} \leftarrow \text{SP}_{[cr:N]} $ \Comment{shift one place for new data}
    		\State $\text{SP}_{[cr]} \leftarrow \text{SP}_{in}$\Comment{insert new SP}
		\Else
    		\State $\text{CR}_{W+1} \leftarrow \text{TSD}_{in}$
    		\State $n_{W+1} \leftarrow 1$
    		\State $\text{SP}_{N+1} \leftarrow \text{SP}_{in}$
		\EndIf
		
		\State \Return $\text{SP}$, $\text{PD}=\{\text{CR}_1:n_1,...\}$, $d'_{max}$ \Comment{RM is $\text{CR}_{i}$ where $\text{SP}_{[n_1+..n_i]} == \text{max}(\text{SP})$}
	\end{algorithmic}
\noindent\makebox[\linewidth]{\rule{0.4 \paperwidth}{0.4pt}}
\end{center}

\section{Simulation studies}
\label{sec:experiments}
The Solar Home data from New South Wales, Australia, is used for the simulation study in this paper, which contains half-hourly PV generation and load demand data of 300 residential consumers~\cite{datasourcewebsite}. Since both the Fixed-memory and codebook-based methods can save memory but are lossy, see Sections~\ref{sec: method-fix} and \ref{sec: method-codebook}, we report and discuss the results of the fixed-memory and codebook models individually to better show their advantages and drawbacks. 
Finally, the overall accuracy of all the proposed methods is compared in Section~\ref{subsec:accuract_all_methods} for different data lengths.

\subsection{Fixed-memory method: different strategies and memory sizes}
\label{exp:memory size}
\begin{figure}[H]
    \centering
    \includegraphics[width=.7 \textwidth]{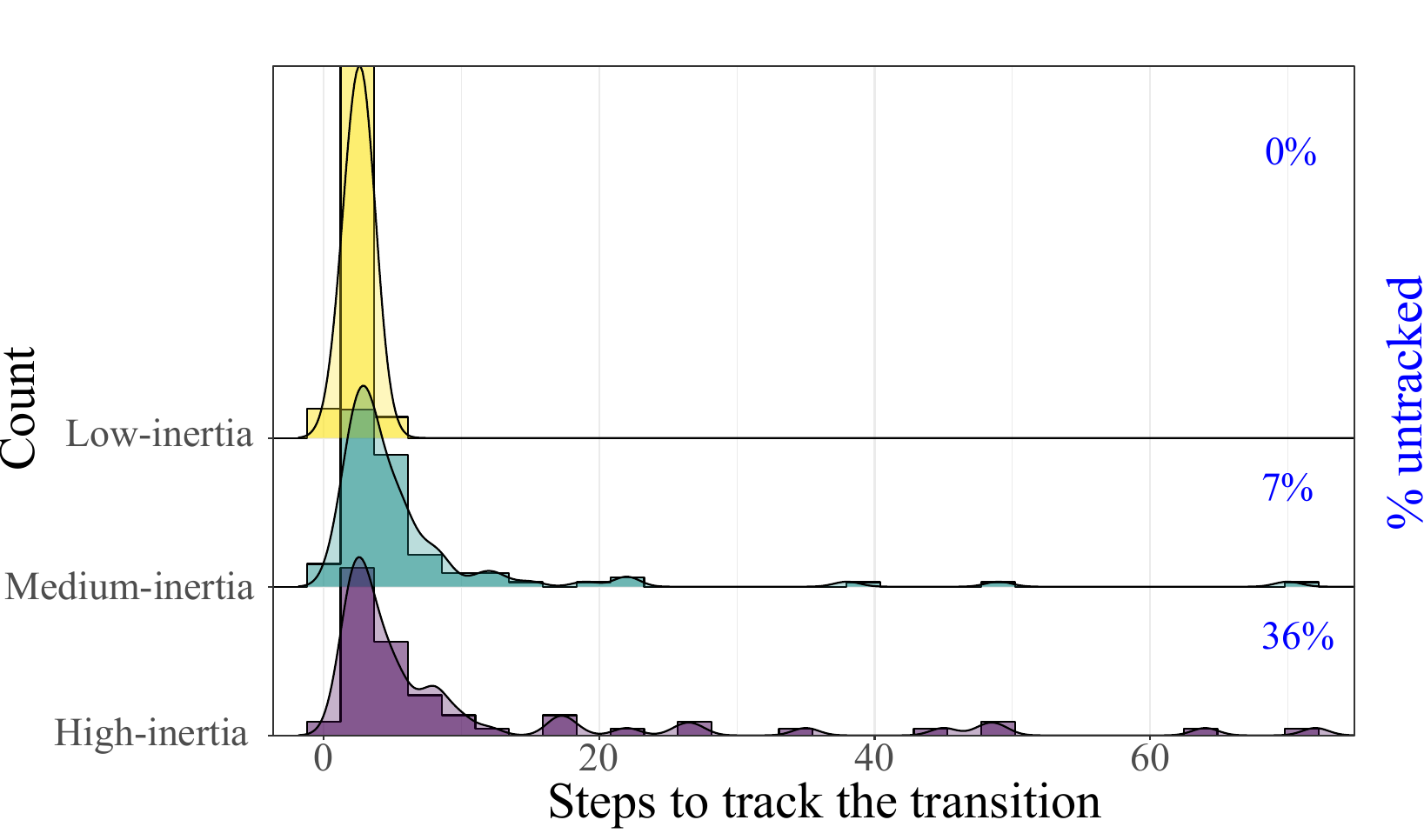}
    \caption{Inertia study for 100 non-solar users becoming solar users: median number of iterations to detect the change is 3, 4 and 11 for low-, medium-, and high-inertia strategies}
    \label{fig:nonsolar-solar-his}
\end{figure}

\begin{figure}[H]
    \centering
    \includegraphics[width=.7 \textwidth]{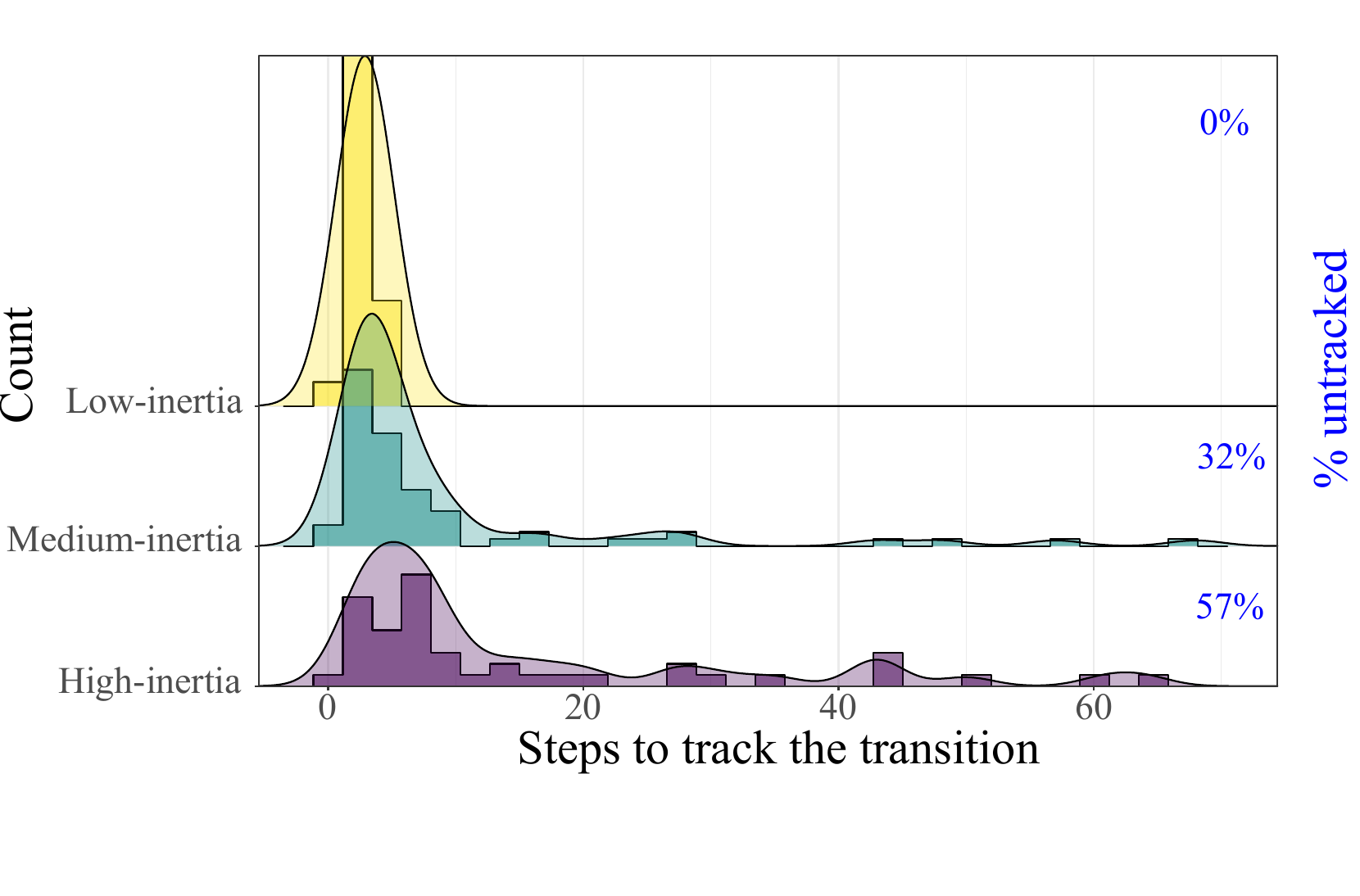}
    \caption{Inertia study for 100 solar users becoming non-solar users: median number of iterations to detect the change is 3, 7 and 80 for low-, medium-, and high-inertia strategies}
    \label{fig:solar-nonsolar-his}
\end{figure}

First, we expanded the illustrative example of Fig.~\ref{fig:nonsolar-solar} to a larger simulation study in which a group of users changed from solar users to non-solar users and vice versa. To do so, 100 users with a rooftop solar system are randomly selected from the dataset. It is assumed that on day 10, their solar system breaks down; hence, they become non-solar users in the remaining days of the simulation study. Similarly, another dataset containing 100 users without a rooftop solar system is created in which they add a solar system on day 10; hence, switching to a solar user in the rest of the simulation study. 
The daily RM updates are recorded to identify the number of updates in each method to detect the switching behaviour.
 Figures~\ref{fig:nonsolar-solar-his} and \ref{fig:solar-nonsolar-his} show the outcomes for solar users becoming non-solar users and vice versa, respectively.




The simulation results reveal that the low-inertia strategy needs the lowest number of days to recognise the change from non-solar to solar users (as shown in Fig.~\ref{fig:nonsolar-solar-his}). Also, 7\% and 36\% of the changes remained unidentified in the medium- and high-inertia strategies, respectively, after 80 days of computations, while the low-inertia strategy had zero unidentified changes during the same time. For detecting solar users becoming non-solar users (as shown in Fig.~\ref{fig:solar-nonsolar-his}), 32\% and 57\% of the changes couldn't be identified by the medium- and high-inertia strategies, respectively, after 80 days, while the low-inertia strategy had zero unidentified cases. The low-inertia strategy generally detects the changes more quickly because the memory contains only the most recent days. The high-inertia strategy, however, keeps the most significant features in the memory; hence, difficult to forget the old patterns. The medium-inertia strategy is balanced compared to the other two. It can track the switching within 30 days for more than half of the users in both experiments. Furthermore, detecting non-solar users changing to solar users is less difficult for both high- and medium-inertia strategies because solar days' sub-patterns provide lower $\text{SP}$ values, which are relatively easier to preserve for the two strategies.

We also ran a simulation study with different memory sizes to assess the impact of memory size on the performance of Fixed-memory approaches, i.e., finding solar users from non-solar users. The performance of the extracted RMs is quantified by the classification accuracy of a single neuron network classifier approach from our previous work~\cite{yuan2022irmac}. The results are reported in Table~\ref{table: fix_window}. The low-inertia strategy's accuracy increases from 82\% to 90.6\% with higher memory sizes, which shows a notable sensitivity to the memory size. This is aligned with our previous discussion on the low-inertia strategy, as it preserves pnly the memory of the latest days. The other two approaches show better accuracy and higher stability than the low-inertia strategy. The high-inertia strategy is the best method to detect long-term behavioural features as it enhances the most significant features while updating RM and is robust to abnormal patterns. 
\begin{table}[H]
    \centering
    \caption{Performance of the three variations of Fixed-memory methods in the solar user identification problem. The highest accuracy value for each row is indicated in bold}
    \label{table: fix_window}
    \resizebox{.65\textwidth}{!}{
    \begin{tabular}{cccc}
        \hline
        \rowcolor{TBLHeader} \backslashbox{Memory size}{Accuracy}& low-inertia & medium-inertia & high-inertia \\
        \hline
         5 days&  82.0\% &92.0\% &\textbf{93.3\%} \\ 
         \hline
         7 days& 85.3\% &91.3\% & \textbf{93.3\%} \\
         \hline
         10 days&  88.0\% & \textbf{94.7\%} &94.0\%\\ 
         \hline
         15 days&  89.0\% & \textbf{92.7\%} &\textbf{92.7\%}\\ 
         \hline
         30 days&  90.6\% & 93.9\% &\textbf{94.7\%}\\ 
         \hline
    \end{tabular}
    }
\end{table}

\subsection{Codebook: Compression rate and accuracy}
\label{exp:compression}
The compression performance is quantified by the \emph{memory saving}, which is the ratio of saved memory versus total memory. Here, the total memory is the one required for the additive method. A sensitivity analysis is conducted for 300 consumers with different lengths of $\text{TSD}$ in summer, namely 1, 2 and 3 months of data. The results are summarised in Table~\ref{tab: compression ratio table}. It can be observed that the memory saving rate increases when including more data. In other words, adding more data may lead to more stable codewords.
\begin{table}[H]
    \centering
    \caption{Compression ratio of the codebook method}
    \label{tab: compression ratio table}
    \resizebox{.5\textwidth}{!}{
    \begin{tabular}{cccc}
        \hline
        \rowcolor{TBLHeader}\backslashbox{Month}{Memory saving}& min & max & average\\
        \hline
         1& 9\% & 96\% & 30\%\\ 
         \hline
         2 & 8\% & 98\% & 31\%\\
         \hline
         3 & 6\% & 98\% & 34\% \\
         \hline
    \end{tabular}
    }
\end{table}

\begin{figure}[H]
    \centering
    \includegraphics[clip, trim=0.1cm .1cm .1cm .1cm,width=.5 \textwidth]{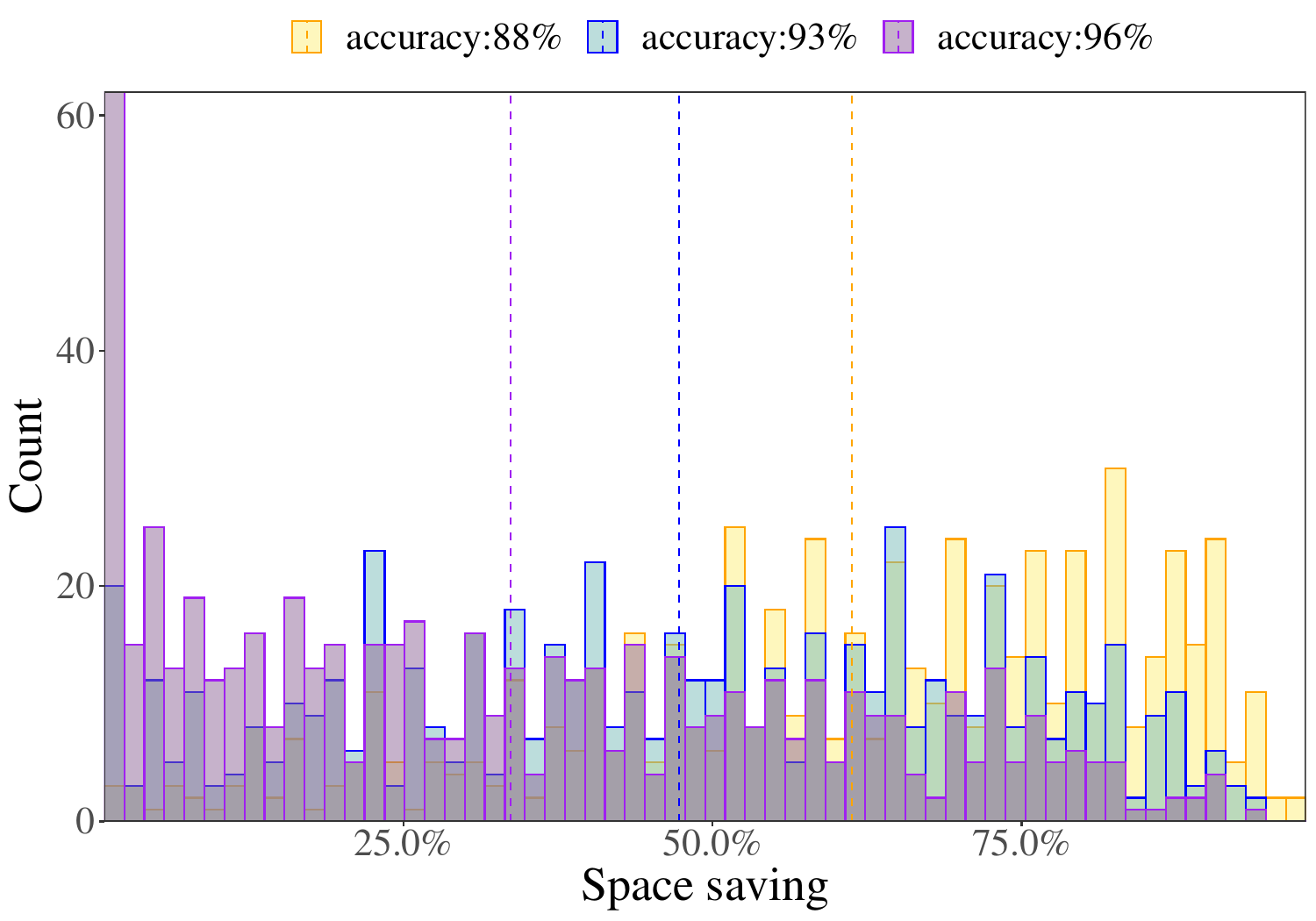}
    \caption{Trade-off between memory saving and accuracy in the codebook-based method for 299 solar users and 299 non-solar users}
    \label{fig:sapce-saving-his}
\end{figure}

One hyperparameter in the codebook-based method is the threshold to determine if one sub-pattern can be replaced with the codeword. A lower threshold results in more codewords in the codebook or PD; hence, less memory saving. On the other hand, the distance between replaced sub-patterns and the codewords will be smaller. As a result, there is a trade-off between memory saving and accuracy. Notably, the compression performance for each user is different because of the variability in their patterns. To measure the trade-off between accuracy and compression, we ran a simulation study in which the threshold is changed from 0.5-2 to compute the memory-saving for the 300 users' data and measure the performance of the extracted RMs by identifying solar users with a linear classifier developed in our previous work~\cite{yuan2022irmac}. The memory-saving distributions for the end-users with different classification accuracy are plotted in Fig.~\ref{fig:sapce-saving-his}.

In general, the users' memory-saving ratio shows high variability, as shown in Fig.~\ref{fig:sapce-saving-his}. At 96\% accuracy, 87 users barely show any memory-saving. However, most users' data can be compressed to some extent, and the average memory saving is 30\%. When the accuracy decreases, it can be seen that the distribution is moving to the higher range of memory saving. At 88\% accuracy on the classification problem, most users achieve more than 50\% memory saving.

\begin{figure}[H]
    \centering
    \includegraphics[width=0.8 \textwidth]{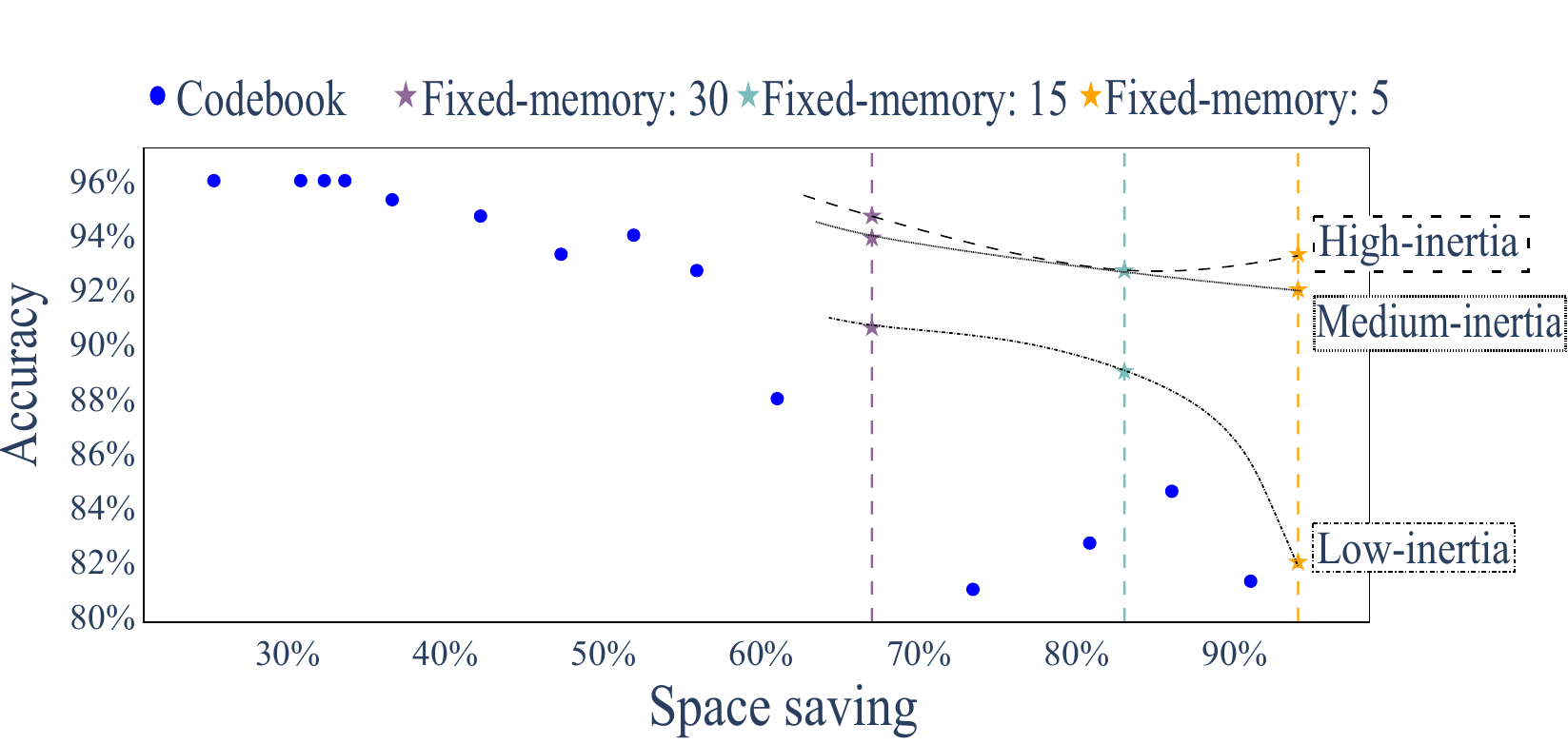}
    \caption{Accuracy vs. compression ratio Pareto front for Codebook and Fixed-memory techniques}
    \label{fig: pareto front}
\end{figure}

A Pareto front is shown in Fig.~\ref{fig: pareto front} by running more experiments on the average memory saving versus accuracy, where three months of data is included for 598 users. The Fixed-memory method with the three elimination strategies is also included for performance comparison. 

It can be seen from Fig.~\ref{fig: pareto front} that the Codebook method without compressed representation achieved $30\%$ memory saving on average without sacrificing accuracy. When more than $50\%$ of memory saving is gained at a higher compression range, the high- and medium-inertia strategies from the Fixed-memory method are significantly better. On the other hand, as we have seen in Table~\ref{tab: compression ratio table}, where the memory-saving has increased with the length of $\text{TSD}$ in the codebook method, this means the accuracy in the high compression range could be higher when there is longer historical data.

\subsection{Accuracy analysis for the proposed updating methods}
\label{subsec:accuract_all_methods}
The overall accuracy and memory size are presented in Table~\ref{tab: overall accuracy}, where the memory size in the Fixed-memory method is 15 (i.e., $M=15$) and the compression rate of the codebook method is $30\%$.

\begin{table*}[ht!]
    \centering
    \caption{Performance of the different methods in the solar user identification problem}
    \label{tab: overall accuracy}
    \resizebox{1\textwidth}{!}{
    \begin{tabular}{cccccc}
        \hline
        \rowcolor{TBLHeader}\backslashbox{Month}{Accuracy} & Additive & Fixed-Memory low& Fixed-Memory high & Fixed-Memory medium & \textbf{Codebook} \\
        \hline
         1& \textbf{94.0\%} & 88.0\% & 92.7\% & 92.0\% &  \textbf{94.0\%}\\ 
         \hline
         2 & 94.7\% & 86.0\% & 94.7\% & \textbf{96.0\%} &  \textbf{96.0\%} \\
         \hline
         3 & \textbf{96.0\%} & 89.3\% & 92.7\% & 92.7\% &  \textbf{96.0\%} \\
         \hline
    \end{tabular}
    }
\end{table*}
The Codebook and Additive methods show the overall best accuracy among the three methods, followed by the Fixed-memory with high-inertia strategy. The accuracy of the Fixed-memory method with a low-inertia strategy significantly varies with different data. 

Based on the simulation studies in previous sections, the main features of the three updating methods are summarised in Table~\ref{tab:pros-cons}.

\begin{table*}[!hb]
\centering
\caption{A comparison of the features of the proposed RM updating method}
\setlength{\extrarowheight}{0pt}
\addtolength{\extrarowheight}{\aboverulesep}
\addtolength{\extrarowheight}{\belowrulesep}
\setlength{\aboverulesep}{0pt}
\setlength{\belowrulesep}{0pt}
\captionsetup{labelformat=empty}
\resizebox{\linewidth}{!}{%
\begin{tabular}{llll} 
\toprule
\rowcolor{TBLHeader} Method & Advantages & Disadvantages & Applications \\\midrule
 \vcell{Additive} & \vcell{\begin{tabular}[b]{@{}>{}l@{}}$\bullet$ Lossless SP values\\$\bullet$ Simple\\$\bullet$ Preserves data\end{tabular}} & \vcell{\begin{tabular}[b]{@{}>{}l@{}}$\bullet$ Memory intensive\\$\bullet$ Highest time complexity\end{tabular}} & \vcell{$\bullet$Long-term behavioural studies with SP} \\[-\rowheight]
\printcelltop & \printcellmiddle & \printcelltop & \printcelltop \\
\hline
 \vcell{Fixed-memory} & \vcell{\begin{tabular}[b]{@{}>{}l@{}}$\bullet$ Lowest memory requirement\\$\bullet$ Fast update (low-inertia)\\$\bullet$ Feature enhancement~(high-inertia)\\$\bullet$ Balanced performance (medium-inertia)\end{tabular}} & \vcell{\begin{tabular}[b]{@{}>{}l@{}}$\bullet$ Low accuracy\\$\bullet$ Permanent knowledge~elimination \\(in low-inertia~strategy)\end{tabular}} & \vcell{\begin{tabular}[b]{@{}>{}l@{}}$\bullet$ Short-term modelling\\$\bullet$ Short-term tracking of changes\end{tabular}} \\[-\rowheight]
 \printcelltop & \printcellmiddle & \printcelltop & \printcelltop \\
\hline
\vcell{Codebook-based} & \vcell{\begin{tabular}[b]{@{}>{}l@{}}$\bullet$ Low memory requirement\\$\bullet$ Preserve the changes~from historical records\\$\bullet$ Flexible\\$\bullet$ High security (no sequential information)\\$\bullet$ High compression ratio~for long TSDs\end{tabular}} & \vcell{\begin{tabular}[b]{@{}>{}l@{}}$\bullet$ Cannot preserve original~data\\$\bullet$ Memory intensive for short~TSDs\end{tabular}} & \vcell{\begin{tabular}[b]{@{}>{}l@{}}$\bullet$ Long-term users\\$\bullet$ Sensitive data to preserve privacy\\{}{}\\{}\end{tabular}} \\[-\rowheight]
 \printcelltop & \printcellmiddle & \printcelltop & \printcelltop \\
\bottomrule
\end{tabular}
}\label{tab:pros-cons}
\end{table*}

\subsection{Future work}

The authors envision three possible directions for future work. Firstly, the memory size for the Fixed-memory method and the pre-defined threshold for the codebook-based method were chosen based on educated guesses. Those two hyperparameters are related to data resolution and applications; hence, more studies can be done on systematical selection of these parameters. Secondly, the extracted RM can be used in more applications, e.g., demand modelling, forecasting, and demand-side management measures. It could be detecting new EV users, new AC/Heatpump installations or shifting between cooling and heating regimes. It might be able to detect malicious behaviours such as electricity theft. This paper uses the solar user identification problem as an example to compare the performance of the extracted RMs for each proposed updating method. More applications can be investigated with the same structure. Thirdly, our study is limited to the utility meters' data accessible to aggregators and utilities. With high-resolution data from smart meters, it is possible to detect the switching behaviours and dis-aggregation problems \cite{Ruzzelli2010, CarrieArmel2013}.  

\section{Conclusion}
\label{sec:conclution}
This paper proposes three RM updating methods and six sub-strategies for electricity residential end-users. The performance and sensitivity of each method to their respective hyperparameters are analysed with a set of simulation studies. Overall, the additive method provides the most accurate results at a high memory and computational power requirement. The Fixed-memory method offers the highest memory saving and a fixed temporal complexity. Three data elimination strategies are proposed that can be used under different circumstances. The low-inertia strategy is sensitive to behavioural changes and memory size; hence, is good for tracking changes. The high-inertia strategy cannot detect sudden changes but is good at enhancing similar sub-patterns in the memory. The medium-inertia strategy is a balanced solution in terms of accuracy, memory-saving, and detecting changes. The Codebook method is the most accurate solution, providing more than 30\% of memory saving without sacrificing accuracy. With this method, both the codebook with compressed representation and PD provides the same RM, but the former can easily recover the whole $\text{TSD}$ while the latter is anonymous. The memory-saving ratio of the codebook method increases with the data size, which makes it a more robust solution for storing long-term records or high-resolution $\text{TSD}$ for end users. Another significant advantage of the Codebook method is that the sub-sequences are stored as codewords; hence, they are more secure at the users' end.


%



\section*{Acknowledgment}

This project is funded jointly by the University of Adelaide industry-PhD
grant scheme and Watts A/S, Denmark.




%
\bibliography{bare_jrnl}

%








\end{document}